\documentclass[12pt,preprint]{aastex}
\usepackage{emulateapj5}

\usepackage{epsfig}
\usepackage{graphicx}
\usepackage{url}
\def\gta{\mathrel{\hbox{\rlap{\hbox{\lower4pt\hbox{$\sim$}}}\hbox{$>$}}}}

\shorttitle{IRAC Zodiacal Light}
\shortauthors{Krick et al.}

\begin{document}

\bibliographystyle{apj}
\title{\bf A Spitzer IRAC Measure of the Zodiacal Light}

\author{Jessica E. Krick \altaffilmark{1}, William J. Glaccum \altaffilmark{1}, Sean J. Carey \altaffilmark{1}, Patrick J. Lowrance \altaffilmark{1}, Jason A. Surace \altaffilmark{1}, James G. Ingalls \altaffilmark{1}, Joseph L. Hora \altaffilmark{2}, William T. Reach\altaffilmark{3}}

\altaffiltext{1}{Spitzer Science Center, MS 220--6,
California Institute of Technology, Jet Propulsion Laboratory,
Pasadena, CA 91125, USA}
\altaffiltext{2}{Harvard-Smithsonian Ctr. for Astrophysics}
\altaffiltext{3}{SOFIA / USRA }
\email{jkrick@caltech.edu}

\begin{abstract}

  The dominant non-instrumental background source for space--based
  infrared observatories is the zodiacal light.  We present $\it
  Spitzer$ Infrared Array Camera (IRAC) measurements of the zodiacal
  light at 3.6, 4.5, 5.8, and 8.0$\;\micron$, taken as part of the
  instrument calibrations. We measure the changing surface brightness
  levels in approximately weekly IRAC observations near the north
  ecliptic pole (NEP) over the period of roughly 8.5 years. This long
  time baseline is crucial for measuring the annual sinusoidal
  variation in the signal levels due to the tilt of the dust disk with
  respect to the ecliptic, which is the true signal of the zodiacal
  light.  This is compared to both Cosmic Background Explorer Diffuse
  Infrared Background Experiment (COBE DIRBE) data and a zodiacal
  light model based thereon.  Our data show a few percent discrepancy
  from the \cite{1998ApJ...508...44K} model including a potential
  warping of the interplanetary dust disk and a previously detected
  overdensity in the dust cloud directly behind the Earth in its
  orbit.  Accurate knowledge of the zodiacal light is important for
  both extragalactic and Galactic astronomy including measurements of
  the cosmic infrared background, absolute measures of extended
  sources, and comparison to extrasolar interplanetary dust models.
  IRAC data can be used to further inform and test future zodiacal
  light models.
\end{abstract}

\keywords{cosmology: diffuse radiation --- interplanetary medium ---
  infrared: diffuse background}

\section{Introduction} 
\label{intro}

The dominant, non-instrumental background source for infrared
observations in space is the zodiacal light (ZL), which comes from both
scattered and thermal interplanetary dust (IPD) emission.  This dust
originates mainly from comets and asteroids, as well as a minimal
amount from the interstellar medium \citep{1996ApJ...472..812G}.  It
has many components, including a smoothly distributed dust cloud
along with various clumps and gaps generated by interactions and
resonances with the large bodies of the solar system.

Studies of the ZL are applicable to at least three
different, unrelated fields of astronomical research.  Understanding
the ZL allows more accurate studies of the cosmic infrared
background (CIB) because it is the dominant error source in those
measurements, which has implications for pop III stars.  In addition,
any absolute measure of extended surface brightness, for example the
intracluster light in galaxy clusters, or the outer expanses of nearby
galaxies, needs to take into account the contribution from zodiacal
light.  Lastly, zodiacal models are important in comparison to
extrasolar IPD models, especially in using structure
in the IPD of stellar systems to find planets.

Measurements of the ZL and models for its properties are
based on IRAS (12 - 100$\;\micron$) and, primarily, COBE DIRBE (1 -
240$\;\micron$)\citep{1992ApJ...397..420B,1993SPIE.2019..180S}. The
model most commonly used to date, based on DIRBE data, is
\cite{1998ApJ...508...44K}.  The DIRBE data have been subsequently
analyzed in conjunction with other data sets to calibrate
inconsistencies in various ZL models.  Motivated by a
desire to measure the CIB, \citet{1998ApJ...496....1W} and
\citet{2000ApJ...536..550G} re-derived a ZL model, based
on DIRBE data only, making changes to the scattering function and
forcing the darkest regions to zero zodiacal emission. Using Infrared
Telescope in Space (IRTS) data, \citet{2005ApJ...626...31M} note the
need for a correction of order a few percent at near-infrared
wavelengths to the \cite{1998ApJ...508...44K} model which they
attribute to calibration differences between COBE and IRTS, and quoted
uncertainties in the model parameters.  \citet{2010A&A...523A..53P}
used a hybrid approach to fix most ZL model parameters to
be consistent with the \cite{1998ApJ...508...44K} model while still
allowing their AKARI Infrared Camera data to constrain certain
parameters.  They find an underestimate by \citet{1998ApJ...508...44K}
of the earth trailing cloud component and a possible warping of the
IPD cloud.  We present here an independent evaluation of the
ZL model of \cite{1998ApJ...508...44K} using $\it Spitzer$
IRAC data \citep{2004ApJS..154....1W, fazio2004}.

$\it Spitzer's$ measurements of the ZL are unique because 1) the
spacecraft is not in the same place as the Earth, which gives us new
positional information on the IPD cloud and 2) we have a multi-year
baseline with approximately weekly cadence.  $\it Spitzer$ is in an
earth-trailing orbit, slowly drifting behind the earth at a rate of
$\sim$0.1 AU per year.  As of January 2012, $\it Spitzer$ is
approximately 1 AU from the earth.  We know its position to 500 km,
considerably under mission specifications. The weekly cadence allows
us to measure the annual variation in the IRAC background levels due
to the ZL signal, which in turn will allow for generation of a more
accurate model than what could be generated from the ground or a
stationary satellite with short duration.

The measured surface brightness of the ZL depends on time of year,
direction of observation, and location within the dust cloud.  The
IRAC wavelength range (3.6 - 8$\;\micron$) probes regimes of both
scattered light and thermal emission; both contribute equally at 3.6,
but the longer wavelengths are dominated by thermal emission
\citep{1994ApJ...431L..63B}. The DIRBE data show that the zodiacal
light typically accounts for just over 50\% of the non-instrumental
measured sky levels at $3.5\;\micron$, which is about five times the
predicted CIB levels.  That number jumps to the zodiacal light being
70\% of the non-instrumental sky levels at $4.5\;\micron$.  At longer
wavelengths the sky levels are a few hundred times brighter than the
CIB signal.  Only redward of $\sim100\;\micron$ do other foregrounds
become significant \citep{1998ApJ...508...25H}.



\section{ Observations \& Data Reduction}

\subsection {Spitzer IRAC}
\label{irac_observations}

As part of the calibration program of IRAC, a field $~3.5^{\circ}$
from the NEP, with relatively low ZL and no bright stars or extended
galaxies is observed with a regular cadence for the purpose of having
a shutterless measurement of the bias and dark current in the arrays
\citep{2009ApJS..185...85K}.  When looking towards the NEP, $\it
Spitzer$ observes a near vertical line of sight through the IPD cloud,
approximately perpendicular to the dust plane, at a distance of about
1AU from the Sun.  During the cryogenic mission from Dec 2003 through
May 2009, dark field data were observed twice per campaign, whenever
the IRAC instrument was on, which was roughly every two to three
weeks.  $\it Spitzer$ cryogen was depleted in May 2009, leaving only
the 3.6 and 4.5$\;\micron$ channels functional.  From May 2009 through
Jan 2010, as the instrument was changing temperature and bias levels,
the dark fields flux levels were not comparable to those at a steady
temperature, and therefore are not used in this analysis.  During the
warm mission, from January 2010 to the present, dark fields are
observed once per week, with relatively few exceptions.

The data used in this analysis are observed as a set of 18, dithered,
100s Fowler-sampled exposures processed with the calibration pipeline
version S18.8(cryo) and S19.0(warm).  The longest possible IRAC
exposure times were chosen as the best measure of the ZL
variations.  For the 8$\;\micron$ data, 50s observations are the longest
possible exposure time.

Similar to the basic calibrated data pipeline detailed in the IRAC
Handbook
\footnote{\url{http://irsa.ipac.caltech.edu/data/SPITZER/docs/irac/iracinstrumenthandbook/}},
each of the raw frames was bias subtracted, linearized, flat-fielded,
and corrected for various instrumental effects.  The bias correction
removes the 'first frame' effect which is a dependency of the bias
pattern on the time since last readout, or 'delay time'. For the
cryogenic mission, a pre-launch library of frames taken at different
exposure delay times was interpolated and used as a first frame
correction.  The warm mission has no such ground data, and therefore
no first frame correction.  However, we do not expect the lack of a
warm first frame correction to effect the ZL measure
because it should be a constant offset in the level of the frames
since all data was taken with the same delay times in both cryogenic
and warm data.  After these corrections, a separate program creates a
median image of all the frames (18) within each set of observations
using outlier rejection, including a sophisticated spatial filtering
stage to reject all stars and galaxies generating a final "skydark"
image.  The calibration pipeline was designed as the best measure of
the dark and bias level in the frames without signal from resolved
sources.


Diffuse stray light is a contaminant at the 1\% level in both the raw
and flat-field images used to correct the pixel to pixel gain effects.
We expect this contaminant to be mitigated to less than 0.25\% by
using 18 dithered frames in the median combine.  All data has been
converted into physical units of MJy/sr by applying a calibration
based on point sources.  Because the 5.8 and 8.0$\;\micron$ detectors
suffer from internal scattering caused by photons diffusing around on
the chip, the point source calibration is not appropriate for extended
sources in these bands.  We therefore apply an extended source
aperture correction which is a $\sim30\%$ increase.  This
correction, which has associated errors, is not critical to the below
analysis because it only effects the overall level of the background,
and not the annual variation.

To measure the background level in the data, which is composed of
ZL, CIB, and any residual instrumental dark current, a
gaussian distribution is fitted to the histogram of all pixel values.
The mean of that gaussian is considered the true background level for
that observation. The mode, as calculated with the technique described
in \citet{bickel2002}, is different from the mean by only 0.35\%,
which is insignificant to our conclusions.  Therefore, we use the mean
as an adequate representation of the background.

Daily observations of calibration stars throughout the mission show
that IRAC photometry in all four channels is stable to 1\%, or better,
as a function of time.  Therefore, it is reasonable to compare
photometry over the entire mission without large time-scale
systematic drifts.

A completely different calibration product exists to measure the flat
fields for IRAC, where observations are taken of high background
ecliptic fields.  Those observations are not a clean ZL
measurement in the same sense as the dark field data because they are
taken of a different field every month, and so they do not have the
same sampling as the dark field and therefore won't have good enough
statistics to compare the sinusoidal variation to the DIRBE
data/models.

\subsection{COBE DIRBE }
\label{dirbe_observations}

We make a direct comparison between IRAC bands 1, 2, 3, \& 4 (3.6,
4.5, 5.8, 8.0$\;\micron$) and COBE DIRBE bands 3, 4, and 5 ( 3.5, 4.9,
and 12$\;\micron$). DIRBE operated in cryogenic mode from November 1989
to September 1990, scanning half the sky every day, building an
all-sky map with high coverage and measuring absolute flux using a
zero-flux internal calibrator.  The DIRBE background has contributions
from ZL, astrophysical sources, and CIB.  For this
analysis we used the calibrated individual observations (CIO), which
are the unbinned data for the full cryogenic mission (285 days), with
data taken every $1/8$
second\footnote{\url{http://lambda.gsfc.nasa.gov/product/cobe/dirbe_exsup.cfm}}.
DIRBE pixels are roughly $20\arcmin$ on a side.  We measured surface
brightness in the single pixel that includes the center of the IRAC
dark field.  At the location of that pixel, over the course of the
cryogenic COBE mission, there are roughly 495 individual observations
per channel after outlier rejection, where $5\%$ of observations were
rejected because they were greater than five sigma from the mean of
the distribution.  We binned the data to give roughly 34 observations.
This binning level was chosen to reduce noise while still retaining
the annual ZL signature.  Errors are fixed per channel at 3.3\%, 2.7\%
and 5.1\% for the three channels respectively
\citep{1998ApJ...508...25H}.

\section{Results}

Figures 1-4 display the IRAC background levels over the entire 8.5
year lifetime, to date, of $\it Spitzer$.  Cryogenic mission data are
shown in green, and warm mission data are shown in blue.  The 5.8 and
8$\;\micron$ channels, shown in Figures 3 and 4, were no longer viable
after cryogenic operations ended $\sim$5.5 years after launch.
Between the cryogenic and warm missions, the detector temperature
increased from 15 to 28.7K, consequently increasing the dark current
in the images by about 0.6MJY/sr.  Therefore, the 3.6$\;\micron$ surface brightnesses in
the cryo and warm data are shown on different scales. The
3.6$\;\micron$ cryogenic data are much noisier than the $3.6\;\micron$
warm data, and all of the 4.5$\;\micron$ data.  The cause of this is
uncertain, but could be that the instrument was less stable due to
constant power cycling and annealing that only occurred during the
cryogenic mission.

The DIRBE data are shown in Figures 1,2,\& 4 as solid black points.
IRAC 8$\;\micron$ data are shown with the DIRBE 12$\;\micron$ channel for
comparison.  The IRAC 5.8 micron data do not have a good counterpart
in the DIRBE filter set.  In all cases the DIRBE data is phased in
time to fit on top of the IRAC observations even though they were not
taken simultaneously.  The DIRBE data are also shifted in surface
brightness to match the IRAC data.  The two data sets do not have the
same absolute level because 1) Filter responses are not the same, 2)
IRAC data includes an instrumental dark current because IRAC does not
have an absolute calibration (see \S\ref{irac_observations}), and 3)
while DIRBE data do have absolute calibration, they include a signal
from stars and galaxies that are unresolved in its relatively large
beam. \citet{2000ApJ...536..550G} measure the contribution of stars to
be roughly 10\% of the total flux at $3.5\;\micron$.


\subsection{Comparison with Zodiacal Light Model}
\label{model}

The \citet{1998ApJ...508...44K} model is based on
DIRBE spectral, temporal and angular information.  Briefly, this is a
complex, 3D, physical model with over 90 free parameters.  It includes
contributions from a smooth cloud, three asteroidal debris dust bands,
and a circumsolar dust ring near 1AU.  Documented sources of uncertainty
include non-uniqueness of the model, use of circular, flat orbits when
ellipticity and warping are known to exist, and simplistic assumptions
about the dust distributions; among others.  We emphasize that this is
an extremely difficult problem to solve with many components and
limited data.

The solid black line in Figures 1-4 shows the predicted ZL level based
on \citet{1998ApJ...508...44K}.  The model values have been shifted in
surface brightness by $0.02$, $0.06$, $-1.77$, and $1.79$ MJy/sr in
the four channels respectively, to match the IRAC mean cryogenic
levels.  The bottom plot of each figure shows the residuals after
subtracting the zodiacal light model from the data.  The y-axis is
shown in units of percent of the surface brightness data.

\section{ Discussion}
\label{discussion}

The sinusoidal variation in our figures is the ZL signature.
We see an anual variation in the ZL contribution to IRAC
data because the location of the dark field precesses around the real
north ecliptic pole over the course of a year, the dust particle
orbits near 1AU are eccentric, and the zodiacal cloud is tilted
relative to the Earth's orbital plane \citep{2010LPI....41.1499R}.
Spitzer is moving above and below the ecliptic dust plane, so that
when the telescope is below the plane, it views a larger column density
of material towards the NEP, and we see a maximum. Six months later,
when the telescope is above the plane, the column towards the NEP is
much smaller and we see a minimum ZL signature.  Because
these data were taken at the NEP, we expect much less ZL
than we would observe edgewise through the ecliptic plane.

The absolute surface brightness measured in IRAC data is affected by
both astronomical sources (zodiacal, interstellar medium, and
extragalactic), and instrumental effects.  While DIRBE made an
absolute measurement of the instrumental background level, IRAC
cannot.  Without use of a shutter, IRAC has no direct way of disentangling
the dark current and bias level from astronomical sources.  We
therefore ignore the absolute level in the plots and focus only on the
sinusoidal shape as the measurement of ZL variation as a function of
time.  The IRAC data reduction process (see \S\ref{irac_observations})
removes all stars and galaxies from the measurement, so the seasonal
variation seen is not the result of the resolved star and galaxy
content changing.  We assume that the ZL is constant over the 25
square arcminute IRAC field of view, and the larger, 1800 square
arcminute, DIRBE beam.

Figures 1-4 show significant residuals between the IRAC data and the
model, implying inaccuracies in the model.  The model curve should fit
the IRAC data to higher precision because it has been tailored to the
IRAC bands, $\it Spitzer$ orbital position around the sun, and the
pointing and time that the data were taken. At 3.6$\;\micron$, the
cryogenic and DIRBE data are too noisy to glean much
information. However the warm mission data show that the model
under-predicts the amplitude of the variation by $\sim 2$ percent.  An
underestimate of the amplitude could imply either an underestimate of
the amount of dust at 1AU, or that the scale height of the dust disk
is flatter than modelled leading to Spitzer traversing further above
and below the concentrated area than predicted.

At 4.5$\;\micron$, all IRAC data show a deviation from a sinusoid shape
in addition to an underestimate of the amplitude of that sinusoid.
The shape change is most clearly evident in the warm mission
residuals, which are not sinusoidal in shape.  This is the only
channel where this shape deviation is seen.  A possible explanation
for the shape change is that the dust disk is warped \citep[which has
been predicted before using IRAS data;][]{1988A&A...196..277D}.  The
4.5$\;\micron$ residuals are larger than the 3.6$\;\micron$ residuals,
which could be giving us color information about the model implying a
better knowledge of the dust grain properties.  However, it could also
be connected to the difference in the scattered and thermal components
of the ZL to the two different wavelength bands, and we
have no way of separating those effects.  The 4.5$\;\micron$ residuals
also show the overdensity behind the earth as seen previously at
8$\;\micron$(see below).

At 5.8$\;\micron$, the data are too noisy to be able to glean
information from the residuals.  Surface brightness values can be
negative because a ground-based estimate of the dark current is
removed from this data, which is known to differ from the true dark
current, of which we have no good measure without use of a shutter. We
can rule out large scale changes from the predicted surface
brightnesses.

At 8.0$\;\micron$, deviations from a simple sinusoidal annual
variation over the first year and a half of the mission have been
associated with the telescope travelling through an overdensity in
dust behind the earth \citep{2010LPI....41.1499R}. This overdensity is
seen in Figure 4 where the residuals are mainly negative for the first
1.5 years, and then switch to oscillating around zero.  Beyond that
overdensity, the residuals look very smooth and constant, implying no
further large over- or under-densities at 1AU.  In addition to the
overdensity, our data show that the model under-predicts the amplitude
of the variation by $\sim5\%$, similar to that seen at 3.6 and
4.5$\;\micron$.


\section{Conclusion}
\label{conclusion}

We used IRAC calibration data taken roughly weekly of the NEP to study
the ZL component at 3.6, 4.5, 5.8, and 8.0$\;\micron$ over
the course of the currently 8.5 year mission of the instrument.  We
compare the IRAC data to both COBE DIRBE data and the ZL
model of \cite{1998ApJ...508...44K} based thereon.  COBE DIRBE data
are taken from 9.4 months of observations of the same region of the
sky as the IRAC data at 3.5, 4.9, and 12$\;\micron$.  The
\cite{1998ApJ...508...44K} model is a 90 parameter fit to the DIRBE
all sky data at multiple wavelengths from 1-240$\;\micron$.  All data
are shown in Figures 1-4.  The sinusoidal variation in the plots is
the ZL signature.  The Spitzer IRAC data show a deviation
from the \cite{1998ApJ...508...44K} model at most at the few percent
level.  We see an under-prediction of the amplitude of the yearly
variation by the model, the presence of an overdensity behind the
earth, and possible evidence for a warping in the IPD cloud.  These
data show both that IRAC can be used for ZL studies and
that the ZL model would benefit from the additional
information gathered here.  A better understanding of the zodiacal
light will have broad impacts on studies of the CIB, low surface
brightness observations, and extrasolar planets, among other things.

 Generating a new ZL model is beyond the scope of this
 work.  It is difficult to know the effect of the few percent
 discrepancies discussed here on work which uses the
 \cite{1998ApJ...508...44K} model.  Because the contribution of the
 ZL to the background of any given image will change
 as a function of direction, time of year, and wavelength, there is no
 easy prescription for the application of these residuals to the
 conclusions of other papers.  It is worth noting that work on the
 cosmic infrared background is very sensitive to models of the
 ZL as that type of science is often working at only the
 few percent level for detections of their signal.


\acknowledgments

We thank the anonymous referee for useful suggestions on the
manuscript.  This research has made use of data from the Infrared
Processing and Analysis Center/California Institute of Technology,
funded by the National Aeronautics and Space Administration and the
National Science Foundation.  This work was based on observations
obtained with the {\it Spitzer} Space Telescope, which is operated by
the Jet Propulsion Laboratory, California Institute of Technology
under a contract with NASA. We acknowledge the use of the Legacy
Archive for Microwave Background Data Analysis (LAMBDA). Support for
LAMBDA is provided by the NASA Office of Space Science.

{\it Facilities:} \facility{Spitzer (IRAC)} \facility{COBE (DIRBE)}



\begin{figure}
\epsscale{0.6}
\plotone{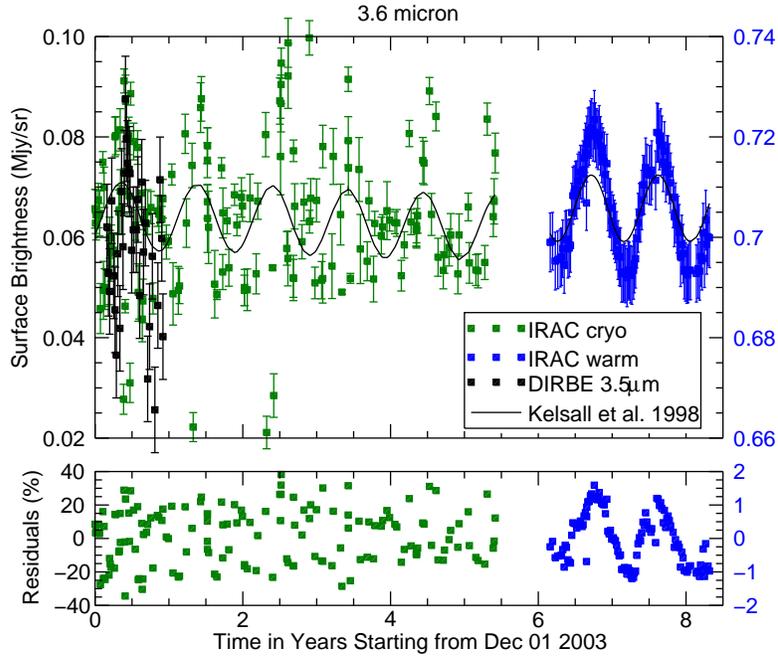}

\caption[ch1]{{\bf Zodiacal light signature.}  Top: IRAC surface
  brightness at 3.6$\;\micron$ of the dark field plotted over the
  timespan in years of the entire Spitzer mission to date. Cryogenic
  data are shown in green with y-axis labels on the left; warm data
  are shown in blue with y-axis labels on the right.  DIRBE data are
  overlaid as the black points.  The \citet{1998ApJ...508...44K} model
  tailored to the IRAC data is shown as the solid black line.  Bottom:
  Residuals in percent after subtracting the
  \cite{1998ApJ...508...44K} models from the IRAC data.  The left
  y-axis corresponds to the cryogenic residuals and the right y-axis
  corresponds to the warm residuals.  Residual levels range from -20
  to 20 kJy/sr for the cryogenic data and -10 to 10 kJy/sr for the
  warm data.}
\label{fig:ch1}
\epsscale{1}
\end{figure}
\begin{figure}
\epsscale{0.6}
\plotone{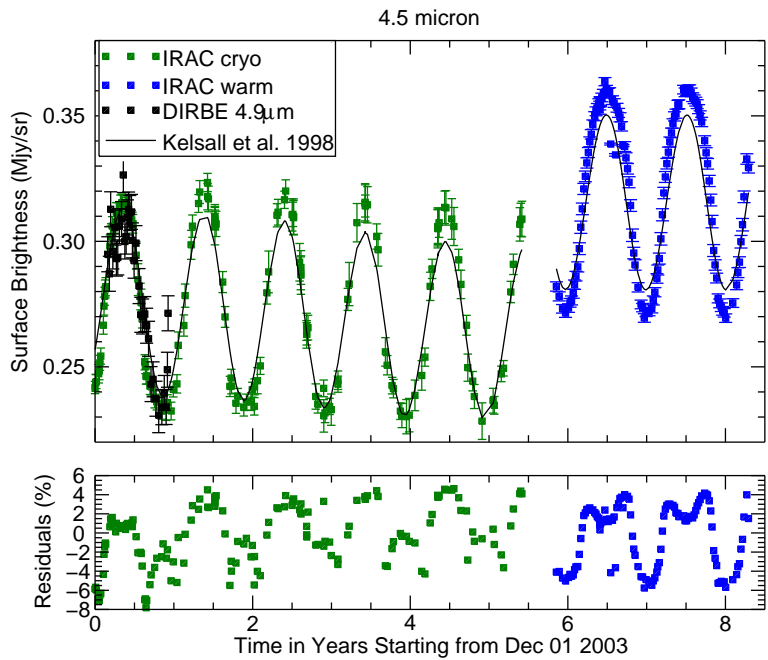}

\caption[ch2]{Same as Figure \ref{fig:ch1}, but for 4.5$\;\micron$
  (IRAC) and 4.9$\;\micron$ (DIRBE).  Residual levels range from -20 to
  15kJy/sr. }
\label{fig:ch2}
\epsscale{1}
\end{figure}
\begin{figure}
\epsscale{0.6}
\plotone{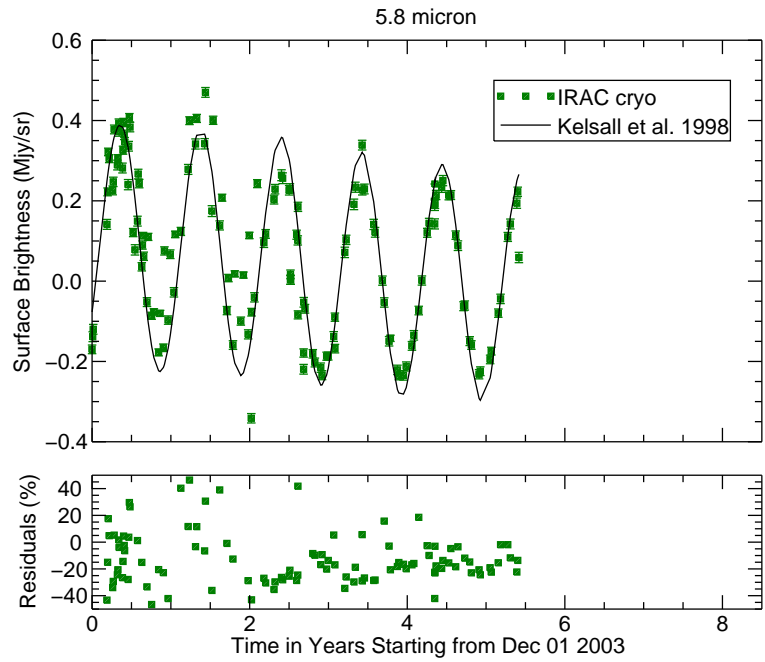}

\caption[ch3]{Same as Figure \ref{fig:ch1} for 5.8$\;\micron$(IRAC).
  The 5.8$\;\micron$ IRAC channel was only usable during the cryogenic
  mission.  Surface brightness levels can be negative due to lack of
  an absolute dark current calibration for IRAC (see
  \S\ref{irac_observations}.  Residual levels range from -100 to 200 kJy/sr.}
\label{fig:ch3}
\epsscale{1}
\end{figure}
\begin{figure}
\epsscale{0.6}
\plotone{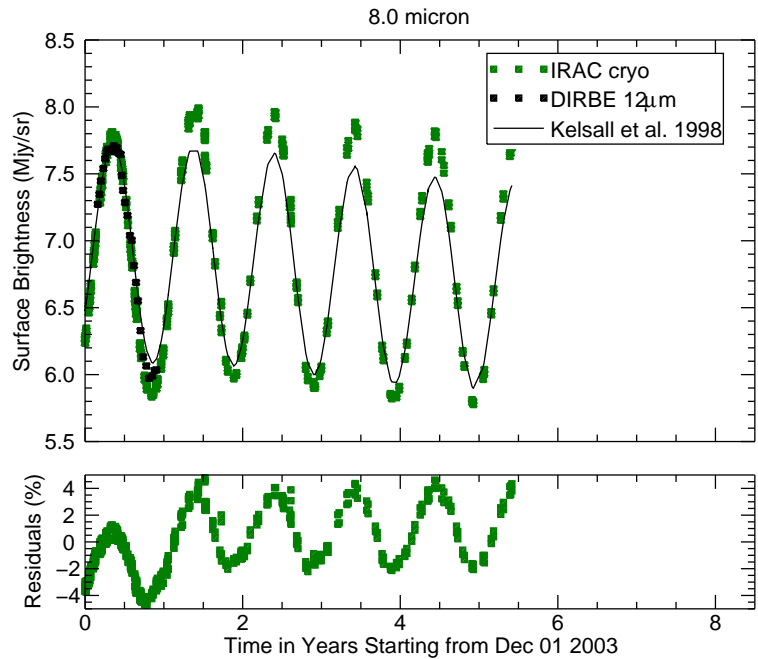}

\caption[ch4]{Same as Figure \ref{fig:ch1} for IRAC 8.0$\;\micron$ and
  DIRBE 12$\;\micron$.  The 8.0$\;\micron$ channel was only usable
  during the cryogenic mission.  Residual levels range from -300 to 400
  kJy/sr.  The negative offset in the residuals for the first 1.5
  years is due to the dust overdensity behind the earth. }
\label{fig:ch4}
\epsscale{1}
\end{figure}


\end{document}